# Interplay of short-range bond order and A-type antiferromagnetic order in metallic triangular lattice GdZn$_3$P$_3$

Jiesen Guo[1], Fan Yang[1], Yuzhou He[2], Xinyang Liu[1,2,3], Zheng Deng[2,4], Qinghua Zhang[2*], Xiancheng Wang[2,4], Xianlei Sheng[1*], Wei Li[3,5,6], Changqing Jin[2,4], and Kan Zhao[1*]

[1]*School of Physics, Beihang University, Beijing 100191, China*

[2]*Beijing National Laboratory for Condensed Matter Physics,*

*Institute of Physics, Chinese Academy of Sciences, Beijing 100190, China*

[3]*CAS Key Laboratory of Theoretical Physics, Institute of Theoretical Physics, Chinese*

*Academy of Sciences, Beijing 100190, China*

[4]*School of Physical Sciences, University of Chinese Academy of Sciences, Beijing 100190, China*

[5]*CAS Center for Excellence in Topological Quantum Computation,*

*University of Chinese Academy of Sciences, Beijng 100190, China*

[6]*Peng Huanwu Collaborative Center for Research and Education,*

*Beihang University, Beijing 100191, China*

*Corresponding author email: zqh@iphy.ac.cn, xlsheng@buaa.edu.cn, kan_zhao@buaa.edu.cn

**Abstract:**

We investigate a hexagonal ScAl$_3$C$_3$-type antiferromagnet GdZn$_3$P$_3$ single crystal. Compared with antiferromagnetic topological material EuM$_2$X$_2$ (M=Zn, Cd; X=P, As), the GdZn$_3$P$_3$ features an additional ZnP$_3$ trigonal planar units. Notably, single-crystal X-ray diffraction analysis reveals that Zn and P atoms within trigonal planar layer exhibit significant anisotropic displacement parameters with a space group of P6$_3$/mmc. Meanwhile, scanning transmission electron microscopy experiment demonstrates the presence of interstitial P atoms above and below the ZnP honeycomb lattice, suggesting potential ZnP bond instability within the ZnP$_3$ trigonal layer. Concerning the triangular Gd$^{3+}$ layer, the magnetic susceptibility $\chi(T)$ and heat capacity measurements reveal long-range antiferromagnetic order at $T_N$ = 4.5 K. Below $T_N$, the in-plane $\chi(T)$ is nearly 4 times the $\chi(T)$ along $c$ axis, indicative of strong magnetic anisotropy. The Curie Weiss fitting to the low temperature $\chi(T)$ data reveals ferromagnetic interaction ($\theta_{cw}$ = 5.2 K) in $ab$-plane, and antiferromagnetic interaction ($\theta_{cw}$ = -1 K) along $c$ axis, suggesting the ground state as an A-type antiferromagnetic order. Correspondingly, the density function theory calculation shows that GdZn$_3$P$_3$ is an indirect semiconductor with a band gap 0.27 eV, supported by the resistivity measurement on polycrystal sample. Interestingly, the GdZn$_3$P$_3$ single crystal exhibits metallic conductivity with an anomaly at $T_N$, likely associated with the observation of interstitial P atoms mentioned above. Therefore, our results establish GdZn$_3$P$_3$ system as a concrete example for investigating the coupling between charge carrier and triangular lattice magnetism in the two-dimensional lattice framework, on the background of short-range bond order.

**Introduction:**

Geometric frustration in magnetism has long been a prominent research focus in condensed matter physics, as this model prevents antiparallel spin alignment in the antiferromagnetic ground state. This results in significant entanglement and fluctuations, giving rise to many novel states, such as spin liquid [1-8], spin ice [9-16] and spin supersolid [17-20]. As the prototype model of the two-dimensional (2D) triangular lattice formed by rare earth ions has attracted more attention, notable examples include the quantum spin liquid candidate YbMgGaO$_4$ [3-8] and ARCh$_2$ (A=alkali or monovalent ions, R=rare earth, and Ch=O, S, Se) [21-24]. While most 2D triangular systems focus on insulators, and the presence of itinerant electrons in the triangular lattice material, the emergence of exotic magnetic order could be driven by the Ruderman–Kittel–Kasuya–Yosida (RKKY) interactions.

Based on the above idea, the hexagonal ScAl$_3$C$_3$-type RM$_3$X$_3$ (R=rare earth; M=Zn, Cd; X=P, As) compounds have been widely studied, with P6$_3$/mmc space group [25-29]. As shown in Fig. 1(a), the crystal structure of RM$_3$X$_3$ features alternating layers of RX$_6$ octahedra, MX$_4$ tetrahedra, and MX$_3$ trigonal planar units, the M$_{trig}$ and X$_{trig}$ atoms of MX$_3$ are cross-arranged to form a honeycomb [26]. To compare, the antiferromagnetic topological materials EuM$_2$X$_2$ (M=Zn, Cd; X=P, As) crystallize in a trigonal system with the P$\bar{3}$m1 (No. 164) space group, constituting by an layered structure composed of edge-sharing MX$_4$ tetrahedra and EuX$_6$ octahedra [30-40]. The interlayer distance between the triangular lattice layer of Eu$^{2+}$&Ce$^{3+}$ ions increases from 7.1683 Å in EuCd$_2$P$_2$ and 7.3278 Å in EuCd$_2$As$_2$, to 10.2387 Å in CeCd$_3$As$_3$, which enhances the two-dimensional character of the RM$_3$X$_3$ structure. Previous studies of CeCd$_3$X$_3$ (X=P, As) have demonstrated strong easy-plane magnetic anisotropy explained by crystalline electric field (CEF) model [41], which demonstrated the low-carrier-density metallic state, with antiferromagnetic (AFM) transition at about 0.4 K [42, 43]. In the J$_{eff}$ =1/2 compound NdCd$_3$P$_3$, the AFM transition temperature of 0.34 K is slightly lower than that of CeCd$_3$X$_3$ (X=P, As) [44], as well as this trend of magnetic transition is similar to that of RZn$_3$P$_3$ (R=rare earth) [28, 45].

During the structural refinement within P6$_3$/mmc space group, both Cd$_{trig}$ and As$_{trig}$ atoms of CeCd$_3$As$_3$ exhibit significantly large displacement parameters [26]. As reported in Ref. [26], this pathological issue can be partially removed by splitting atom positions, or incorporating twins in a low-symmetry space group. Fig.1(a) illustrates an acceptable model that the Cd$_{trig}$ and As$_{trig}$ atoms are shifted from the ideal 2c and 2d sites (with -6m2 symmetry) to 6h sites (mm2 symmetry) at a partial occupancy of 1/3. Within the 6h sites, the 1/3 occupancy indicates the Cd and As atoms are randomly occupied in the three positions, forming a local distorted arrangement within the CdAs$_3$ trigonal planar layer. The scenario could be treated as a dynamic movement of Cd$_{trig}$ acting on the surrounding As$_{trig}$ atoms, which may cause the bond instability of (CdAs)$_{trig}$.

Recently, Ref. [46] provides direct evidence of the local distortion of CdP bonds associated with one-dimensional CdP chains in RCd$_3$P$_3$ (R=La, Ce, Pr, and Nd) single crystals, based on the observation of diffuse intensity in synchrotron X-ray diffraction measurement. By mapping the Cd-P dimer covering problem onto a 2D Ising model defined on the dual triangular lattice, the effective interactions associated with the local bond order on the honeycomb lattice hinder the development of long-range order [46]. Therefore, the coexistence of bond instability and frustrated magnetism renders the RCd$_3$P$_3$ system particularly intriguing, as it extends the concept of geometrical frustration from solely encompassing the magnetic component of electrons to incorporating both spin and charge degrees of freedom. It is crucial to identify an isostructural compound with a relatively high magnetic transition temperature in order to facilitate the investigation of the interplay between these two frustrated subsystems.

In this paper, we present the synthesis and characterization of iso-structural GdZn$_3$P$_3$ single crystal. The ZnP bond instability within ZnP$_3$ trigonal planar units has been revealed through both single crystal X-ray diffraction and scanning transmission electron microscopy (STEM) experiment. Based on the magnetic and thermodynamic data, we identify the ground state of GdZn$_3$P$_3$ as an A-type AFM order below $T_N$ = 4.5 K. The GdZn$_3$P$_3$ system is calculated to be a semiconductor with a band gap of 0.27 eV by DFT method, consistent with the value of 0.2 eV extracted from polycrystalline resistivity data. During single crystal growth, the presence of interstitial P atoms would cause hole-doping into the GdZn$_3$P$_3$ system, inducing the metallic conductivity with an anomaly at $T_N$. Thus, the metallic GdZn$_3$P$_3$ single crystal gives rise to the unique opportunity of investigating the emergent phenomenon in antiferromagnetic RM$_3$X$_3$ system, such as the topological hall effect related with the unconventional spin texture, as well as the charge carrier induced superconductivity.

**Experimental:**
The polycrystals of CeCd$_3$As$_3$ and GdZn$_3$P$_3$ were synthesized by solid-state method. The CdAs, CdP, and ZnP precursors were first weighed in a glove box using a stoichiometric ratio of Cd powder (99.99%), As lump (99.99%), P lump (99.99%) and Zn powder (99.99%), then mixed in an agate mortar and sealed in an evacuated quartz tube. CdAs was sintered at 500 °C for 30 h, and CdP and ZnP were sintered at 250 °C for 30 h. CdAs and Ce (99.99%), CdP and Gd (99.99%), and ZnP and Gd were mixed at stoichiometric ratios and sintered. As shown in Fig. S1-S2 [47], the phase pure CeCd$_3$As$_3$ and GdZn$_3$P$_3$ polycrystals have been successfully obtained under sintering at 700 °C for 30 h and 1100 °C for 30 h. High-quality single crystals of GdZn$_3$P$_3$ were grown by 0.5 g of stoichiometric amounts of Gd lump (99.99%), Zn powder (99.99%), and P lump (99.99%) together with 4 g of the NaCl/KCl (1:1) flux [25, 28]. The sample was annealed for 24 h at 500 °C, followed by 170 h at 900 °C, then furnace cool to room temperature. The plate-like single crystals were collected by washing with deionized water.

X-ray diffraction (XRD) patterns were measured using a Bruker D8 ADVANCE diffractometer with Cu Kα (λ=1.5418 Å) radiation at room temperature. The single-crystal X-ray diffraction (SCXRD) data were performed using a Rigaku XtaLAB Synergy diffractometer at room temperature with the Mo Kα (λ=0.71073 Å) radiation. The single crystal structure solution was completed using Olex2 [48] and SHELXL [49]. The high-angle annular-dark-field (HAADF) image and annular bright-field (ABF) image were characterized using an ARM-200F (JEOL, Tokyo, Japan) scanning transmission electron microscope (STEM) operated at 200 keV with a CEOS Cs corrector (CEOS GmbH, Heidelberg, Germany) to cope with the probe-forming objective spherical aberration. The magnetic properties were measured using the Quantum Design Magnetic Property Measurement System (MPMS). The specific heat and resistivity were measured using the Quantum Design Physical Property Measurement System (PPMS). The resistivity was also measured using the Cryogen Free Measurement System (CFMS).

Our DFT uses the generalized gradient approximation (GGA) [50, 51] in the form of PBE functional [52], and included spin-orbit coupling, as implemented in the Vienna ab initio Simulation Package (VASP) [53-55]. The energy cutoff of the plane-wave basis is set to 350 eV. The energy convergence criterion in the self-consistent calculations is set to $10^{-6}$ eV. A 9×9×3 Γ-centered Monkhort-Pack k-point mesh is used for the first Brillouin zone sampling. To account for the correlation effects for Gd, we adopted the GGA + U method with the value of U=1 eV.

**Results and Discussion:**

**A. Crystal structure: X-ray diffraction and Scanning transmission electron microscope**

A series of (0, 0, 2L) peaks are observed by XRD diffraction experiment in Fig. 1(b), indicating the *ab* plane as the natural cleavage plane. Rocking curve analysis of the Bragg peak (008) in the inset of Fig. 1(b) demonstrates a narrow full-width-at-half-maximum (FWHM) of 0.17°, indicating high quality of the $GdZn_3P_3$ crystals. The single crystal diffraction spots along [001] in Fig. 1(d) show direct evidence for the six-degree symmetry of hexagonal $GdZn_3P_3$ system, with related spots along [100] and [010] shown in Fig. S4 [47]. The Rietveld refinements of the SCXRD data converge to the agreement factors *$R_1$=3.61%* and *$wR_2$= 8.96%* in Fig. 1(c), with *a = b* = 4.0107(3) Å and *c* = 19.8808(15) Å in Table 1. And the corresponding crystal structure is depicted in Fig. 2(c), with the refined atomic parameters listed in Table 2.

Due to the presence of large voids above and below the location in Fig. 2(c), $Zn_{trig}$ atom exhibits large anisotropic displacement parameters, with $U_{33}$ = 0.048 along *c*-axis and $U_{11}$ = 0.032 in *ab*-plane at room temperature. These values are comparable to the values of 0.0501 and 0.02826 for $Cd_{trig}$ atom of $PrCd_3P_3$. On the other hand, the $P_{trig}$ atom has larger in-plane displacement with $U_{11}$ parameter as 0.031, also comparable with the value of 0.0392 for $P_{trig}$ atom of $PrCd_3P_3$. The relatively large anisotropy parameters of $Zn_{tirg}$ and $P_{trig}$ for $GdZn_3P_3$ suggest a tendency toward local frustrated bond order within the ZnP honeycomb plane, as observed in isostructural $PrCd_3P_3$ [46].

To reveal the dynamic disorder of interlayer atoms in atomic scale, we continue to perform STEM on a *bc*-plane of $GdZn_3P_3$ about 4 μm × 3 μm area, with details information in Fig.S5 [47, 56]. As shown in Fig.2(a), the selected area electron diffraction (SAED) pattern contains a series of diffraction peaks, such as (001), (002), (003), and (004), etc., and (0-10). Moreover, the HAADF image in Fig. 2(b) directly shows the atomic arrangement in the unit cell of $GdZn_3P_3$ projected along [100] direction, corresponding to the crystal structure in Fig. 2(c). First, the HAADF image confirms the full occupancy of $Zn_{trig}$ and $P_{trig}$ atom within the $ZnP_3$ trigonal planar. We conclude that the underlying bond instability of $(ZnP)_{trig}$ are local characteristics rather than uniform.

Notably, numerous interstitial P atoms are observed in the HAADF image, as indicated by the yellow spheres in Fig. 2(b). Within the honeycomb lattice of $Zn_{trig}$ and $P_{trig}$ atoms, the bond distance is about 2.32 Å; while the distance between $Zn_{trig}$ and $P_{tet}$ atom is 3.29 Å along *c* axis. Thus, the presence of interstitial P atom is closely related with the large voids above and below the location of $Zn_{trig}$ atoms in Fig. 2(c). And we will discuss the effect of interstitial P atoms in electronic transport section later.

## B. Magnetic Properties and Magnetic Structure

Magnetic susceptibility $\chi(T)$ measurements of GdZn$_3$P$_3$ have been performed under magnetic field along $a$, $a^*$, and $c$ axis, showing a clear AFM transition at $T_N$ = 4.5 K, as shown in Fig. 3(a-b) and Fig. S6-7 [47]. The $\chi(T)$ curve exhibits no significant difference between $a$ and $a^*$ directions, with its value being about 4 times than that along $c$ axis below $T_N$. In Fig. S7, The Curie-Weiss law is used to fit the $\chi(T)$ (200-300 K) under 0.05 T yield $\mu_{eff}$ = 7.93 $\mu_B$/Gd in $ab$-plane, consistent with the value 7.94 $\mu_B$. In Fig. 3(a-b), the Curie-Weiss law is used to fit the $\chi(T)$ (15-25 K), which yields $\theta_{cw}$ = 5.2 K for $H//a^*$, indicating ferromagnetic (FM) interaction dominates in the $ab$-plane, and $\theta_{cw}$ = -1 K for $H//c$, indicating AFM interaction dominates along $c$ axis case. These characteristics suggest the formation of an A-type AFM order, likely with the spin alignment of Gd$^{3+}$ ions lying in the $ab$-plane below $T_N$ and FM layers stacked antiferromagnetically.

Similarly, the A-type AFM order has also been observed in topological material EuM$_2$X$_2$ (M=Zn, Cd; X=P, As), with $T_N$ at about 23 K in EuZn$_2$P$_2$ [35-37, 39, 40] and 9 K in EuCd$_2$As$_2$. The much lower $T_N$ in GdZn$_3$P$_3$ is speculated to be largely due to the extra honeycomb nets "ZnP" layer between triangular lattice of Gd atoms, thus weakening the AFM interaction strength. According to the previous literature, the magnetic moments of EuCd$_2$As$_2$ and EuZn$_2$As$_2$ are determined to be parallel to $ab$-plane through resonant elastic x-ray scattering (REXS) and neutron diffraction; and the spins of EuZn$_2$P$_2$ are canted out of the a-a plane by an angle of about 40°± 10° and aligned along the [100] direction as confirmed by resonant magnetic x-ray diffraction data [30, 34, 35, 37].

We also note a subtle cusp appears in $ab$-plane $\chi(T)$ curve at 2.4 K (see Fig.3(a) and Fig. S6), and to capture this feature, Fig. S6(a) displays the temperature derivative of $\chi(T)$ ($d\chi/dT$) as a function of temperature. A modulated magnetic order observed in GdV$_6$Sn$_6$ indicates the presence of an incommensurate magnetic order below 5.2 K and a commensurate magnetic order below 3.8 K, as demonstrated by REXS measurements [57, 58]. To clarify the microscopic magnetic structure together with its temperature dependence of GdZn$_3$P$_3$, it seems necessary to conduct REXS and neutron scattering measurements in future.

In addition, the $\chi(T)$ curves at different magnetic fields are measured as shown in Fig. 3(c-d). With the magnetic field increased to 0.6 T for $H//a^*$ and 1.4 T for $H//c$, the AFM transition peak has been completely suppressed to below 1.8 K for $H//a^*$ and 2.5 K for $H//c$. The $\chi(T)$ curves as temperature derivatives for $H//a^*$ and $H//c$ at various magnetic fields are presented (Fig. S8) [47]. As shown in Fig. 3(e-f), the experiment saturated magnetization $M_S$~6.7 $\mu_B$ for $H//a^*$ and $M_S$~6.5 $\mu_B$ for $H//c$ at 1.8 K, consistent with the value 7 $\mu_B$ for Gd$^{3+}$ ions. From the corresponding $dM/dH$ in Fig. 3(e), these transition behaviors were observed at 0.6 T at 1.8 K and 0.5 T at 3 K under field along $a^*$ axis. Finally, Fig.4 (a-b) depicts the $H-T$ phase diagram constructed by combining the AFM transition peaks under different magnetic fields.

## C. Magnetic heat capacity and entropy

Fig. 5 presents the temperature dependence of measured heat capacity $C_p$ curve, which reveals a sharp transition close to $T_N$ = 4.5 K. Above 30 K, the background specific heat of phonon contribution is fitted by $C_{ph}=aT+bT^3+cT^5+\cdots$ [59]. The magnetic heat capacity $C_m$ is obtained after subtracting phonon contribution. There is still residual magnetic entropy below 1.8 K, so the lower temperature magnetic heat capacity is fitted through the power law $C_m(T)=b\times T^a$ to get a=1.16 as shown in Fig. 5(b), which does not follow the $C_m(T)\sim T^3$ relation observed in the conventional ungapped AFM magnon excitations [60]. By integrating $(C_p-C_{ph})/T$ from 15 K to 0 K, the obtained values of the magnetic entropy $S_{mag}$ are 16.3 J/mol·K, which is 94% of the theoretically maximum magnetic entropy as $Rln8$. The hear capacity $C_p$ data below 1.8 K seems necessary to obtain a more accurate power law fitting as well as the integrated $S_{mag}$ value.

### D. Electronic band structure DFT calculation and Transport results

Similar to EuCd$_2$As$_2$ case [33], GdZn$_3$P$_3$ may have three possible A-type AFM structures as shown in Fig. 6(a) with magnetic moments of Gd$^{3+}$ ions along *b*, *c,* and *x* directions. Consistent with the magnetic evidence above, the DFT calculations verify the configuration with the moment aligned along *b* axis exhibits the lowest free energy. The ground state would be most likely the A-type magnetic structure with Gd moment along *b* axis. Fig.6(b) and Fig.S9 display the corresponding band structures with spin-orbit coupling included and Density of states (DOS) plot near the Fermi level, with indirect band gaps from Γ to M as 0.27 eV, 0.28 eV, and 0.26 eV, respectively [47]. This band gap is smaller than that of polycrystalline LaCd$_3$P$_3$ (0.75 eV) and NdCd$_3$P$_3$ (0.63 eV). Besides, the indirect band gap of EuZn$_2$P$_2$ is calculated to be 0.2 eV, and observed as 0.09 eV in the angle-resolved photoemission spectroscopy (ARPES) experiment [37].

Fig. 6(c) illustrates the electrical resistivity of polycrystalline GdZn$_3$P$_3$ within the temperature range of 2-300 K, consistent with the semiconducting behavior observed in RCd$_3$P$_3$ (R=La, Ce, and Nd) polycrystals. As depicted in inset of Fig.6(c), the transport gap of GdZn$_3$P$_3$ is extracted from a fit to the resistivity data between 250 and 300 K using the thermally activated Arrhenius form $\rho = \rho_0 e^{E_{ac}/2k_BT}$, yielding a band gap of approximately 0.2 eV, consistent with the DFT calculation results above. Interestingly, the AFM transition also shows up in the resistivity curve, marked as $T_N$ = 4.5 K in Fig. 6(c). A similar characteristic has also been observed in EuZn$_2$P$_2$ and EuCd$_2$As$_2$ [21, 51].

As shown in inset of Fig.6(d), the magnetoresistance (MR) curve below 20 K exhibits an initial increase under small field (H < 0.3T), which may be associated with the impurity carriers, similar behavior also reported in EuZn$_2$As$_2$ [38]. Under high field, the magnetic scattering would be suppressed in polarized FM state, thus a negative MR has been observed in polycrystalline GdZn$_3$P$_3$ under field up to 12 T. According to Fig.3(e-f), the GdZn$_3$P$_3$ compound enters the polarized state under 1 T for H//$a^*$ and 4 T for H//$c$, Thus, the negative MR below 5K quickly reaches a magnitude of -60% under 4 T, then saturates at this value under H > 4T. The magnitude of negative MR in GdZn$_3$P$_3$ is compared to that of EuZn$_2$As$_2$ [38], EuZn$_2$P$_2$ and EuCd$_2$As$_2$ [21, 49]. The MR curve becomes more pronounced as temperature decreases from 20 K to 2 K, indicating the significant influence of magnetic ordering behavior on the electronic transport properties in GdZn$_3$P$_3$.

Fig. 6(e) illustrates the resistivity curve of GdZn$_3$P$_3$ single crystal between 2-35 K, exhibiting metallic behavior with an AFM transition peak around $T_N$ = 4.5 K. The resistivity value herein is on the order of 1 mΩ·cm, smaller than the value of 10 mΩ·cm, reported for GdZn$_3$P$_3$ single crystal in Ref. [28]. The emergence of metallic conductivity seems closely related with the presence of interstitial P atoms above and below the ZnP honeycomb lattice, revealed by the STEM experiment mentioned above. And the different resistivity value might be caused by the different density of interstitial P atom in this study. Through the RKKY exchange interaction, the charge carrier could also influence the magnetic property. Correspondingly, the value of $\chi(T)$ here is about 2 times than that in Ref. [28] under H//*ab* plane below $T_N$ = 4.5 K.

**Discussion and Conclusion:**

For the magnetic topological system EuMnSb$_2$ [61-63] and EuCd$_2$As$_2$ [32], both materials exhibit narrow-gap semiconductor behavior. The presence of vacancies at Eu sites can introduce hole-type charge carriers and lead to metallic conductivity within the system. Concurrently, the enhanced RKKY interaction between the Eu magnetic moments and these charge carriers slightly alter the temperature-dependent magnetic ordering behavior, resulting in modifications to the spin configuration in the ground-state.

According to ref. [44], the RCd$_3$P$_3$ would be semiconductor with band gap between 0.6-0.8 eV, verified by the DFT calculation and the polycrystal results. Given the narrow band gap of the RCd$_3$P$_3$ family, these compounds are likely dopable into a metallic state as well. During previous research, single crystals of RCd$_3$X$_3$ (R=La, Ce; X=P, As) exhibits metallic behavior, with the resistivity on the order of 10 mΩ·cm (1 mΩ·cm) for P (As) compounds [42, 43]. Based on the Hall effect measurements, the hole-type carrier density in CeCd$_3$P$_3$ (CeCd$_3$As$_3$) is estimated to be 0.002 (0.003) carriers per formula unit (f.u.) [42, 43]; however, the exact origin remains unclear.

Here, the ZnP bond instability in GdZn$_3$P$_3$ has been demonstrated through both X-ray diffraction and STEM measurement. Our atomic resolution STEM measurement reveals the presence of interstitial P atoms close to the ZnP honeycomb lattice, which would cause hole-doping into the GdZn$_3$P$_3$ single crystal. Our investigation gives an important insight into the metallic behavior of RZn$_3$P$_3$ and RCd$_3$X$_3$ (R=La, Ce; X=P, As) single crystals, thus it would be interesting to conduct the STEM measurement on the above single crystal to reveal its origin in atomic scale.

As mentioned above, the local bond instability in ZnP layer could influence the rare-earth magnetic ion triangular layer through the charge carrier mediated RKKY magnetic interaction as observed in GdZn$_3$P$_3$; meanwhile, the influence could also be realized through the crystal electronic field (CEF) effect for rare-earth magnetic ion with orbital moment such as in RZn$_3$P$_3$ (R=Ce, Pr, Nd, and Sm) system, which deserves further investigation in future. In summary, given the relatively high AFM transition at $T_N$ = 4.5 K, the metallic GdZn$_3$P$_3$ single crystal opens the possibility of investigating the interplay between short-range bond order and triangular lattice magnetism in two-dimensional RM$_3$X$_3$ compounds.

**Acknowledgements**

The authors would like to thank Peijie Sun, Junsen Xiang, Jiabing Xiang, Huifen Ren, Tao Sun and Shaokui Su for helpful discussions and experimental support. The work was supported by the Beijing Natural Science Foundation (Grant No. JQ24012), National Key R&D Program of China (Grant No. 2023YFA1406003, 2022YFA1402600, 2023YFA1406300 and 2024YFA1409500), National Natural Science Foundation of China (Grants No. 12274015, 12174018 and 52322212), and the Fundamental Research Funds for the Central Universities. The authors acknowledge the facilities, and the scientific and technical assistance of the Analysis & Testing Center, Beihang University. A portion of this work was carried out at the Synergetic Extreme Condition User Facility (SECUF).

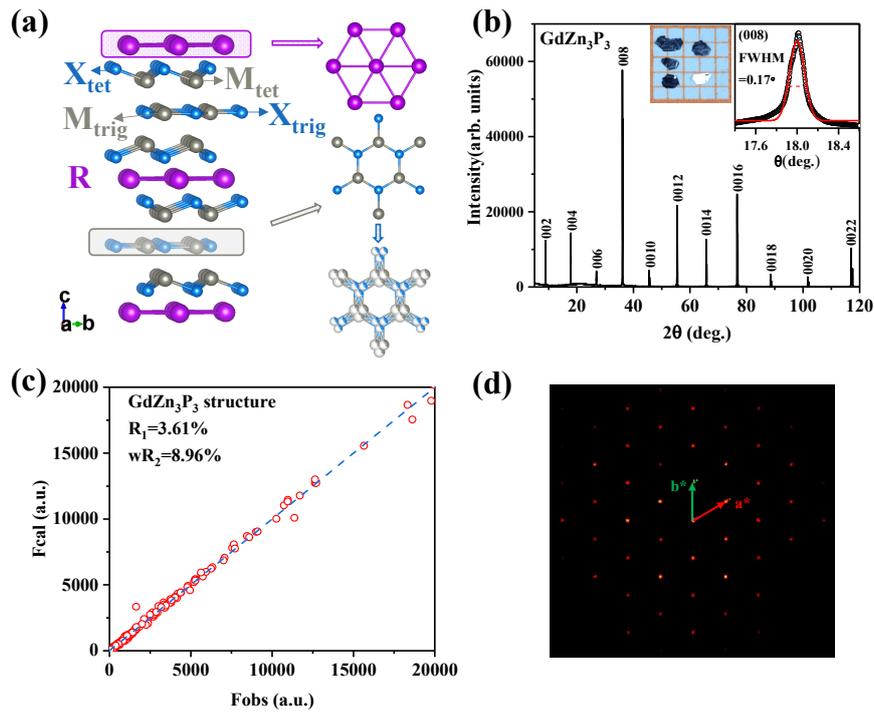

Fig. 1. (a) The crystal structure of RM$_3$X$_3$ (R=rare earth; M=Zn, Cd; X=P, As), with space group P6$_3$/mmc. Purple, grey, and blue spheres represent R, M, and X atoms, respectively. The triangular layer formed by R ions. The honeycomb network formed by MX$_3$ trigonal planar units, which M$_{trig}$ and X$_{trig}$ occupy 1, to the local distorted arrangement in which M$_{trig}$ and X$_{trig}$ are randomly distributed over three equivalent positions, respectively. (b) The XRD pattern of the GdZn$_3$P$_3$ single crystal, with the inset illustrating the Rocking curve scan of the (008) peak fitted by Gaussian function, and a photograph of the single crystals also included. (c) Plots of calculated vs experimental structure factors for the refined structure at room temperature. (d) The diffraction spot pattern taken looking along the [001] from the single-crystal X-ray diffraction of GdZn$_3$P$_3$.

Table 1. Crystal data and structure refinement
for single-crystal GdZn$_3$P$_3$.

| Formula | GdZn$_3$P$_3$ |
|---|---|
| T(K) | 293(2) |
| Crystal system | Hexagonal |
| Space group | P6$_3$/mmc |
| a(Å) | 4.0107(3) |
| b(Å) | 4.0107(3) |
| c(Å) | 19.8808(15) |
| V(Å)$^3$ | 276.95(4) |
| θ | 2.049-28.158 |
| No. reflections collected | 1066 |
| No. of variables | 14 |
| Final R indices | R$_1$=0.0361, wR$_2$=0.0896 |
| R indices (all data) | R$_1$=0.0405, wR$_2$=0.0911 |
| Goodness of fit | 0.992 |
| Largest diff. peak and hole(eÅ$^{-3}$) | 0.92, -1.09 |

Table 2. Wyckoff positions, atomic coordinates, occupancies, isotropic and anisotropic
displacement parameters (Å$^2$), where $U_{23}=U_{13}=0$ for GdZn$_3$P$_3$.

| Atom. | Wyck. | x | y | z | Occu. | U(eq) | U$_{11}$ (U$_{22}$) | U$_{33}$ | U$_{12}$ |
|---|---|---|---|---|---|---|---|---|---|
| Gd | 2a | 0 | 0 | 0.5 | 1 | 0.0095(5) | 0.0075(6) | 0.0134(8) | 0.0038(3) |
| Zn$_{tet}$ | 4f | -1/3 | 1/3 | 0.6299(1) | 1 | 0.0136(7) | 0.0139(9) | 0.0130(11) | 0.0069(5) |
| Zn$_{trig}$ | 2d | -2/3 | 2/3 | 3/4 | 1 | 0.0374(13) | 0.0321(19) | 0.048(3) | 0.0160(9) |
| P$_{tet}$ | 4f | 2/3 | 1/3 | 0.4155(3) | 1 | 0.0112(12) | 0.0117(17) | 0.010(2) | 0.0058(8) |
| P$_{trig}$ | 2c | -1/3 | 1/3 | 3/4 | 1 | 0.024(2) | 0.031(4) | 0.011(4) | 0.0153(18) |

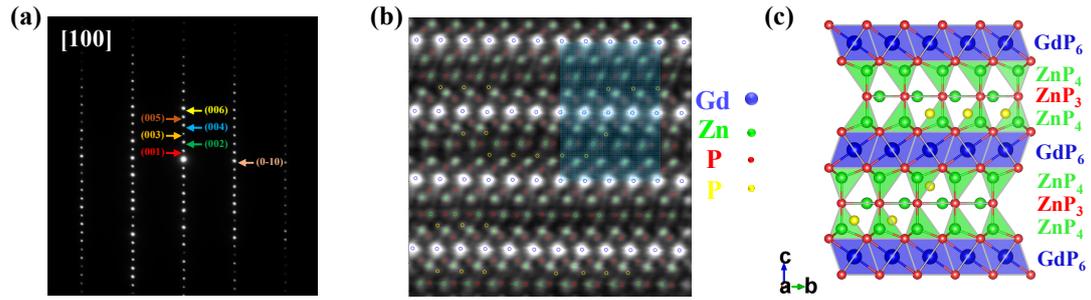

Fig. 2. (a) The electron diffraction pattern along [100] direction. (b) The high-angle annular dark-field (HAADF) image of GdZn$_3$P$_3$ taken along [100] direction. The blue, green, red, and yellow spheres represent Gd, Zn, P, and the interstitial impurity P, respectively. (c) Corresponding crystal structure of GdZn$_3$P$_3$ along [100] direction, matching the blue-boxed region in Fig.2(b). The structure features alternating layers of GdP$_6$ octahedra, ZnP$_4$ tetrahedra, and ZnP$_3$ trigonal planar units, with the red ball marks the anticipated site of the interstitial P atom.

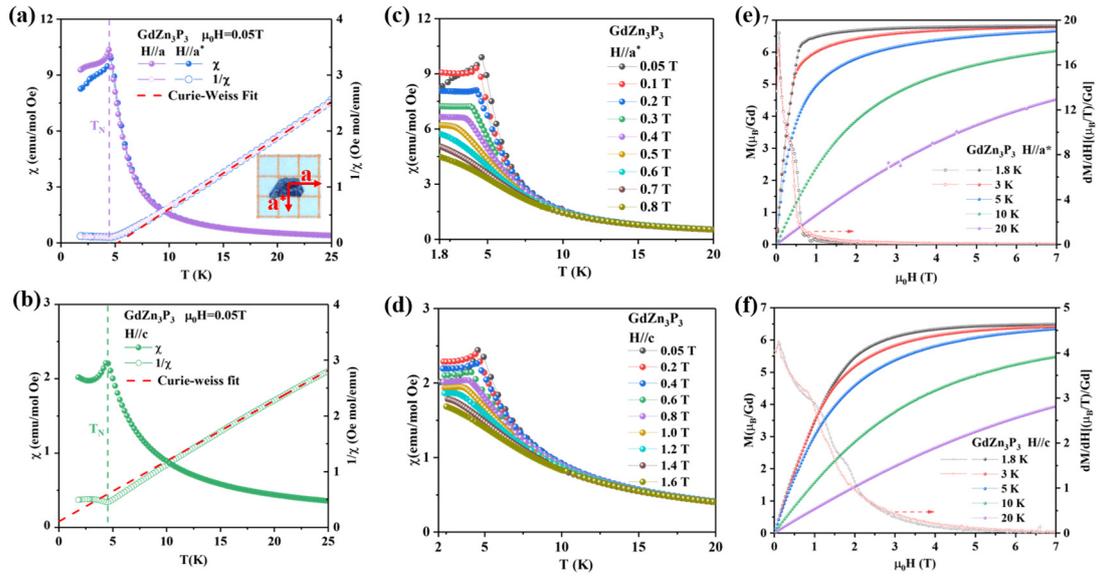

Fig. 3. (a, b) The temperature-dependent magnetic susceptibility χ(T) of GdZn$_3$P$_3$ between 1.8 and 25 K measured with magnetic field along a, a* and c directions, an inset optical image of the single crystal. (c, d) The χ(T) below 20 K at various fields along $a^*$ and $c$ directions. (e, f) Isothermal magnetization and the differential susceptibility dM/dH at different temperatures with the magnetic field applied along the $a^*$ and $c$ directions.

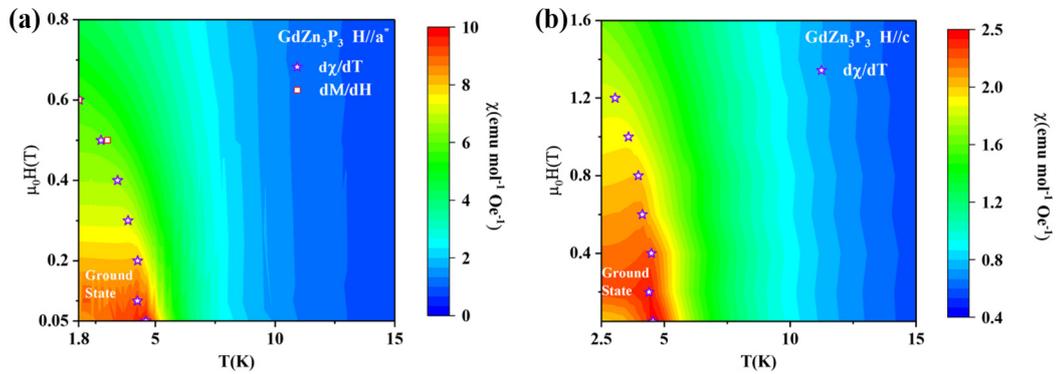

Fig. 4. (a, b) The H-T magnetic phase diagram overlaid on the contour plots of the magnetic susceptibility of GdZn$_3$P$_3$ with the field along the a* and c directions, respectively. The magnetic phase boundaries were extracted through χ(T) and M(H) measurements. The star and square symbols respectively denote the temperature at a specific field where these transition peaks observed in the temperature derivative of susceptibility dχ/dT and the differential susceptibility dM/dH.

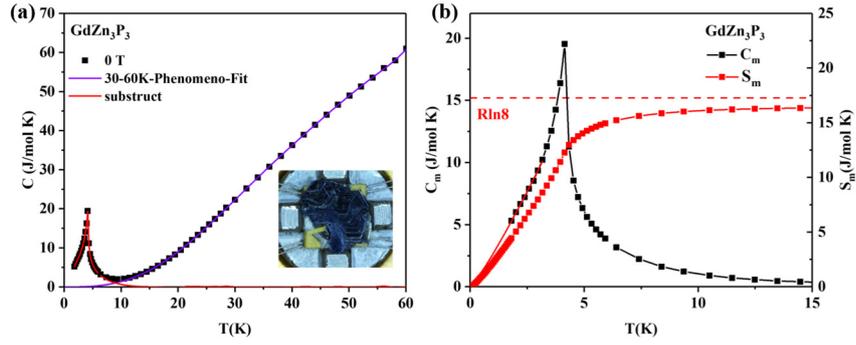

Fig. 5. (a) Temperature-dependent Specific-heat (black square) of GdZn$_3$P$_3$ in zero field, temperature dependence of the magnetic specific heat C$_m$(T) (red line) with the phonon contribution (purple line) removed, and an inset photo of the single crystal for heat capacity measurement in the PPMS puck. (b) The C$_m$ (black line) with the portion of the zero field C$_m$(T) fitted to a b×T$^a$ power law (red extension line) and temperature-dependent integrated magnetic entropy (red line).

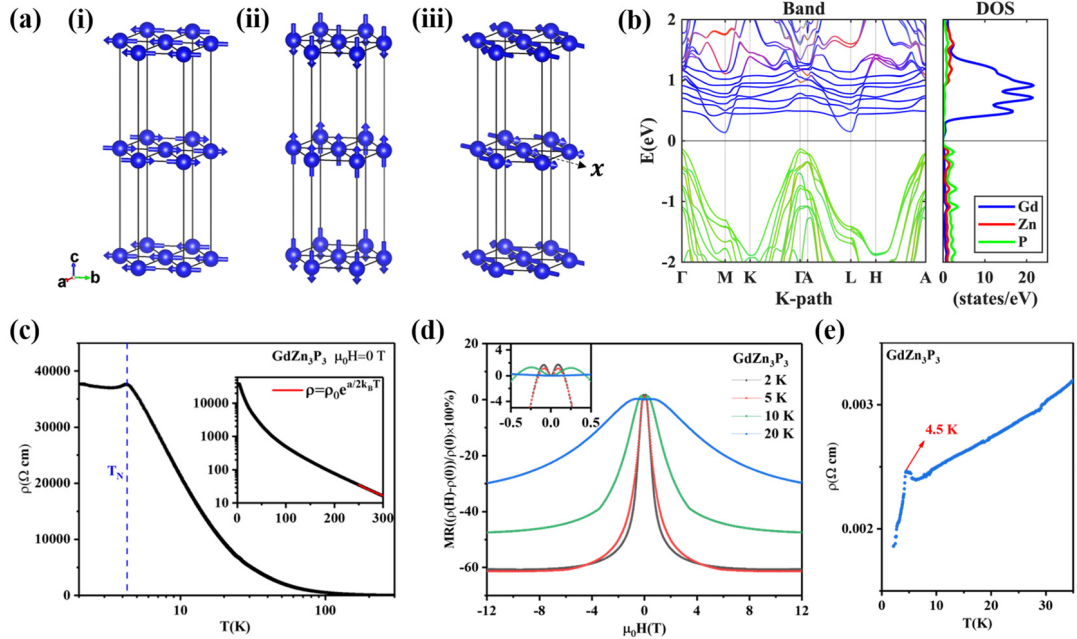

Fig. 6. (a) Proposed possible magnetic structure of GdZn$_3$P$_3$. A-type AFM structure configurations on Gd sites with the magnetic moment aligned along $b$ (i), $c$ (ii) or $x$ (iii) directions. (b) (i) corresponding Band structure with spin-orbit coupling included and Density of states (DOS) plot near the Fermi level, calculated via Density functional theory (DFT). (c) The temperature-dependent resistivity of polycrystalline GdZn$_3$P$_3$ sample between 2 K and 300 K, with the inset shows the transport activation gap extracted from a fit in the 250-300 K range. (d) Magnetoresistance (MR) of GdZn$_3$P$_3$ as a function of field at several representative temperatures, with the inset reveals the detailed behavior of MR in the field range of –0.5 T to 0.5 T. (e)The temperature-dependent resistivity of single crystal GdZn$_3$P$_3$ between 1.8 K and 35 K.